\documentclass[aps,prd,twocolumn,superscriptaddress,floats,amsfonts,amsmath,amssymb]{revtex4}
\usepackage{graphicx,epsf}
\usepackage[active]{srcltx}
\usepackage{bm}
\usepackage[]{latexsym}

\def\bi#1{{\bm #1}}

\newcommand{\sfrac}[2]{\raisebox{0.095ex}{\scriptsize${\frac{#1}{#2}}$}}%

\newcommand{\bea}{\begin{eqnarray}}\newcommand{\eea}{\end{eqnarray}}
\newcommand{\be}{\begin{eqnarray}}\newcommand{\ee}{\end{eqnarray}}
\newcommand{\ba}{\begin{array}}\newcommand{\ea}{\end{array}}
\newcommand{\bit}{\begin{itemize}}\newcommand{\eit}{\end{itemize}}
\newcommand{\ben}{\begin{enumerate}}\newcommand{\een}{\end{enumerate}}

\newcommand{\lab}{\label}

\newcommand{\lf}{\left}

\newcommand{\non}{\nonumber}
\newcommand{\ran}{\rangle}

\newcommand{\ri}{\right}

\newcommand{\om}{\omega}

\begin{document}

\title{Uncertainty relation on world crystal
and its applications to micro black holes}

\author{Petr~Jizba}
\email{jizba@physik.fu-berlin.de}
\affiliation{ITP, Freie Universit\"{a}t Berlin, Arnimallee 14
D-14195 Berlin, Germany} \affiliation{FNSPE, Czech Technical
University in Prague, B\u{r}ehov\'{a} 7, 115 19 Praha 1, Czech Republic}
%
\author{Hagen~Kleinert}
\email{kleinert@physik.fu-berlin.de}
\affiliation{ITP, Freie Universit\"{a}t Berlin, Arnimallee 14
D-14195 Berlin, Germany}
\author{Fabio~Scardigli}
\email{fabio@phys.ntu.edu.tw}
\affiliation{Leung Center for Cosmology and Particle Astrophysics (LeCosPA),
Department of Physics, National Taiwan University, Taipei 106,
Taiwan}

\begin{abstract}

We formulate generalized uncertainty
relations in a crystal-like universe whose lattice spacing is
of the order of Planck length
--- ``world crystal''. In the particular case when energies lie near
the border of the Brillouin zone, i.e., for Planckian energies, the
uncertainty relation for position and momenta does not pose any
lower bound on involved uncertainties. We apply our results to micro
black holes physics, where we derive a new mass-temperature relation
for Schwarzschild micro black holes.  In contrast to standard
results based on Heisenberg and stringy uncertainty relations, our
mass-temperature formula predicts both a finite Hawking's
temperature and a zero rest-mass remnant at the end of the micro
black hole evaporation. We also briefly mention some connections of
the world crystal paradigm with 't~Hooft's quantization and double
special relativity.

\vspace{4mm}

\end{abstract}

\maketitle



\section{Introduction}

Recent advances in gravitational and quantum physics indicate that
in order to reconcile the two fields with each other, a dramatic
conceptual shift is required in our understanding of spacetime. In
particular, the notion of spacetime as a continuum may need revision
at scales where gravitational and electro-weak interactions become
comparable in strength~\cite{Witten:96}. For this reason there has
been a recent revival of interest in approximating the spacetime
with discrete coarse-grained structures at small, typically
Planckian, length scales. Such structures are inherent in many
models of quantum-gravity, such as spacetime foam~\cite{garay:98},
loop quantum gravity~\cite{rovelli:04,gambini:99,alfaro:00},
non-commutative
geometry~\cite{camelia:00,susskind:00,camelia:01,douglas:01},
black-hole physics~\cite{jacobson:08} or cosmic cellular
automata~\cite{Toffoli,Wolfram1,Wolfram2,thooft1,thooft2}.

Despite a vast gap between the Planck length ($\ell_p \approx
1.6\cdot10^{-35}$ m) and smallest length scales that can be probed
with particle accelerators ($\approx 10^{-18}$ m), the issue of
Planckian physics might not be so speculative as it seems.
In fact, probes such as Planck Surveyor~\cite{PS} or
the related
IceCube~\cite{IC}--- which just started or are planned to start
in the near future, are supposed to set various important limits
on prospective models of the Planckian world.

One of the simplest toy-model systems for Planckian physics is
undoubtedly a discrete lattice. Discrete lattices are routinely
used, for instance, in computational quantum field
theory~\cite{Creutz:83,Klein1,FIRST}, but with a few notable
exceptions~\cite{Friedberg:83a,Friedberg:83b,MVF}, they mainly serve
as numerical regulators of ultraviolet divergences. Indeed, a major
point of renormalized theories is precisely to extract
lattice-independent data from numerical computations. One may,
however, investigate the consequences of taking
the lattice no longer as a mere computational device, but as a {\em
bona-fide} discrete network, whose links define the only possible
propagation directions for signals carrying the interactions between
fields sitting on the nodes of the network.

Recently one of us proposed a model of a discrete, crystal-like
universe --- ``world crystal"~\cite{FIRST,MVF,rema0}. There, the
geometry of Einstein and Einstein-Cartan spaces can be considered as
being a manifestation of the defect structure of a crystal whose
lattice spacing is of the order of $\ell_p$. Curvature is due to
rotational defects, torsion due to translational defects. The elastic
deformations do not alter the defect structure, i.e., the geometry
is invariant under elastic deformations. If one assumes these to be
controlled by a second-gradient elastic action, the forces between
local rotational defects, i.e., between curvature singularities, are
the same as in Einstein's theory~\cite{ANN}. Moreover, the elastic
fluctuations of the displacement fields possess logarithmic
correlation functions
 at long distances, so that the memory of the crystalline structure
is lost over large distances. In other words, the Bragg peaks of the
world crystal are not $\delta $-function-like, but display the
typical behavior of a quasi-long-range order, similar to the order
in a Kosterlitz-Thousless transition in two-dimensional
superfluids~\cite{MVF}.

The purpose of this note is to study the generalized uncertainty
principle (GUP) associated with the quantum physics on the world
crystal and to derive physical consequences related to micro black
hole physics. In view of the fact that
micro black holes might be formed at energies as low as the
TeV range~\cite{Dimopoulos:98,Barrau:04,Giddings:08} --- which will be shortly
available in particle accelerators
such as the LHC, it is hoped that the presented
results may be more than of a mere academic interest.

The structure of our paper is as follows: In Section~\ref{SecII}  we
present some fundamentals of a differential calculus on a lattice
that will be needed in the text. In Section~\ref{Sec.III} we
construct position and momentum operators on a 1D lattice and
compute their commutator. We then demonstrate that the usual
Weyl-Heisenberg algebra $W_1$ for $\hat{p}$ and $\hat{x}$ operators
is on a 1D lattice deformed to the Euclidean algebra $E(2)$. By
identifying the measure of uncertainty with a standard deviation we
derive the related GUP on a lattice. This is done in
Section~\ref{Sec.IV}. There we focus on two critical regimes:
long-wave regime and the regime where momenta are at the border of
the first Brillouin zone. Interestingly enough, our GUP implies that
quantum physics of the world-crystal universe becomes
``deterministic" for energies near the border of the Brillouin zone.
In view of applications to micro black hole physics, we derive in
Section~\ref{SecV} the  energy-position GUP for a photon.
Implications for micro black holes physics are discussed in
Section~\ref{SecVI}. There we derive a mass-temperature relation for
Schwarzschild micro black holes.  On the phenomenological side, the
latter provides a nice resolution of a long-standing puzzle: the
final Hawking temperature of a decaying micro black hole remains
{\em finite}, in contrast to the infinite temperature of the
standard result where Heisenberg's uncertainty principle operates.
Besides, the final mass of the evaporation process is zero, thus
avoiding the problems caused by the existence of massive black hole
remnants. Entropy and heat capacity are discussed in
Section~\ref{entr}. Finally, in Section~\ref{appl} we outline a
connection of our results with 't Hooft's approach to deterministic
quantum mechanics and with deformed (or double) special relativity.
Section~\ref{dics2a} is devoted to concluding remarks. For
completeness, we present in Appendix an alternative derivation of
the micro black hole mass-temperature formula.

\section{Differential calculus on a lattice\label{SecII}}

In this section we quickly review some features of a differential
calculus on a 1D lattice. An overview discussing more aspects of
such a calculus can be found,
e.g., in Refs.~\cite{Creutz:83,Klein1,FIRST}. Independent and very
elegant derivation of these can be also done in the framework of a
non-commutative geometry~\cite{Con1,Dim1,Dim2}.

On a lattice of spacing $\epsilon $ in one dimension, the lattice sites
lie at $x_n=n \epsilon $ where $n$ runs through all integer numbers.
There are two fundamental derivatives of a function $f(x)$:
\begin{eqnarray}
(\nabla f)(x) &=& \frac{1}{\epsilon} \left[
f(x + \epsilon) - f(x) \right]\, , \nonumber \\
(\bar{\nabla} f)(x) &=& \frac{1}{\epsilon}
\left[ f(x) - f(x - \epsilon) \right]\, .
\end{eqnarray}
They obey the generalized Leibnitz rule
\begin{eqnarray}
({\nabla} fg)(x) \ &=& \
({\nabla}f)(x)
g(x) + f(x + \epsilon ) ({\nabla} g)(x)\, ,
\nonumber \\
(\bar{\nabla} fg)(x) \ &=& \
(\bar{\nabla} f)(x) g(x) +
f(x-\epsilon)
(\bar{\nabla} g)(x)\, .
\end{eqnarray}
On a lattice, integration is performed as a summation:
\begin{equation}
\int dx \ \!f(x) \ \equiv \ \epsilon \sum_x f(x)\, ,
\label{1@}\end{equation}
where $x$ runs over all $x_n$.

For periodic functions on the lattice or for functions vanishing at
the boundary of the world crystal, the lattice derivatives can be
subjected to the lattice version of integration by parts:
\begin{eqnarray}
&&\sum_x f(x)\nabla g(x)\ = \ - \sum_x g(x)\bar\nabla f(x)\,
,\label{PAR}\\~~~~~ &&\sum_x f(x)\bar\nabla g(x) \ = \ - \sum_x g(x)\nabla
f(x)\, . \label{2@}\end{eqnarray}
One can also define the lattice Laplacian as
\begin{equation}
\nabla\bar \nabla f(x)= \bar \nabla \nabla f(x)= \frac{1}{ \epsilon
^2}[ f(x+ \epsilon )-2 f(x)+ f(x- \epsilon )]\, , \label{LAST}
\end{equation}
which reduces in the continuum limit to an ordinary Laplace operator
$\partial ^2_{x}$. Note that the lattice Laplacian can also be
expressed in terms of the difference  of the two lattice
derivatives:
\begin{equation}
\nabla\bar \nabla f(x) \ =  \ \frac{1}{ \epsilon }
\left[
\nabla f(x)-
\bar\nabla f(x)
\right]\, .
\label{4@}\end{equation}
The above calculus can be easily extended to any number $D$ of
dimensions~\cite{Creutz:83,Klein1,MVF}.
%

\section{Position and momentum operators on a lattice\label{Sec.III}}

Consider now the quantum mechanics (QM) on a 1D lattice in a
Schr\"{o}dinger-like picture. Wave function are square-integrable
complex functions on the lattice,  where ``integration" means here
summation, and scalar products are defined by
\begin{equation}
\langle f|g\rangle \ = \ \epsilon \sum_x f^*(x)
g(x)\, .
\label{int5}
\end{equation}
It follows from Eq.~(\ref{PAR}) that
\begin{eqnarray}
\langle f|\nabla  g \rangle \ = \ -\langle \bar{\nabla} f| g \rangle \, ,
\label{int6}
\end{eqnarray}
so that $(i\nabla)^\dag = i\bar{\nabla} $, and neither $i\nabla$
nor $i\bar{\nabla}$ are hermitian operators. The lattice Laplacian
(\ref{LAST}), however, is hermitian.

The position operator $\hat{X}_{\epsilon}$ acting on wave functions
of $x$ is defined by a simple multiplication with $x$:
\begin{equation}
(\hat{X}_{\epsilon} f)(x) \ = \ x f(x)\, .
\end{equation}
Similarly we can define  the lattice momentum operator
$\hat{P}_{\epsilon}$. In order to ensure hermiticity we should relate
it to the symmetric lattice derivative~\cite{Dim1,Vit1,Klein1}.
Using (\ref{int6}) we have
\begin{eqnarray}
(\hat{P}_{\epsilon}f)(x) \ &=& \ \frac{\hbar}{2i}
[(\nabla f)(x) +
(\bar{\nabla}f)(x)]\nonumber \\
&=&\  \frac{\hbar}{2i\epsilon} [f(x + \epsilon) - f(x -
\epsilon)]\, .
\label{p_eps}
\end{eqnarray}
For small $\epsilon $, this reduces to the ordinary momentum
operator $\hat p\equiv -i\hbar \partial _x$, or more precisely
\be \hat{P}_\epsilon \ = \ \hat{p} + {\cal{O}}(\epsilon^2)\, .
\label{II.26a} \ee
The ``canonical'' commutator between $\hat{X}_{\epsilon}$ and
$\hat{P}_{\epsilon}$ on the lattice
reads
\begin{eqnarray}
\left( [\hat{X}_{\epsilon}, \hat{P}_{\epsilon}] f\right)(x) \ &=& \
\frac{i\hbar}{2}[f(x+\epsilon) + f(x-\epsilon)]\nonumber \\
&\equiv& \ i\hbar(\hat{{I}}_{\epsilon}f)(x)\, .
\label{CR}
\end{eqnarray}
The last line defines a lattice-version of the unit operator as the
average over the two neighboring sites. Note that all three operators
$\hat{X}_{\epsilon}, \hat{P}_{\epsilon}$, and $\hat{{I}}_{\epsilon}$
are hermitian under the scalar product (\ref{int5}).

It was noted in~\cite{Vit1} that the operators $\hat{X}_{\epsilon}$,
$\hat{P}_{\epsilon}$ and $\hat I_{\epsilon}$ generate the Euclidean
algebra $E(2)$ in 2D. Indeed, setting $\hat{M} = \epsilon
\hat{X}_{\epsilon}$, $\hat{P}_1 = \epsilon \hat{P}_{\epsilon}/\hbar$ and
$\hat{P}_2 = \hat I_{\epsilon}$ we obtain
\begin{displaymath}
[\hat{M},\hat{P}_1]\ = \ i\hat{P}_2\, , \;\;\; [\hat{M},\hat{P}_2]\ = \ -i\hat{P}_1\, , \;\;\;
[\hat{P}_1,\hat{P}_2]\ = \ 0\, .
\end{displaymath}
The generator $\hat{M}$ corresponds to a rotation, while $\hat{P}_1$
and $\hat{P}_2$ represent two translations. In  the limit $\epsilon
\rightarrow 0$, the Lie algebra of $E(2)$ contracts to the standard
Weyl-Heisenberg algebra $W_1$: $\hat X_\epsilon\rightarrow \hat x$, $\hat
{P}_\epsilon\rightarrow \hat p$, $\hat {I}_\epsilon\rightarrow
\hat{\openone}$. Thus  ordinary QM is obtained from
lattice QM by a contraction of the $E(2)$ algebra, with the lattice spacing $\epsilon$
playing the role of the {\em deformation parameter}.

All functions on the lattice can be Fourier-decomposed with wave
numbers in the Brillouin zone:
\begin{equation}
f(x) \ =  \ \int_{-\pi/\epsilon}^{\pi/\epsilon}
\frac{dk}{2\pi} \ \! \tilde{f}(k)e^{ikx}\, ,
\label{FT1}
\end{equation}
with the coefficients
\begin{equation}
\tilde{f}(k) \ = \ \epsilon \sum_{x} f(x) e^{-ikx}\, .
\label{FT2}
\end{equation}
This implies the good-old de Broglie relation
\begin{equation}
(\hat p \tilde f)(k)\ = \ \hbar k   \tilde f(k)\, , \label{DEB}
\end{equation}
and its lattice version
\begin{equation}
(-i\nabla \tilde f)(k)\ = \ K  \tilde f(k), ~~~ (-i\bar\nabla \tilde
f)(k)\ = \ \bar K  \tilde f(k)\, , \label{DEB2}
\end{equation}
with the eigenvalues
\begin{equation}
K \ \equiv \  (e^{ik \epsilon }-1)/i \epsilon \ = \ \bar K^*\, .
\label{@}
\end{equation}
From (\ref{DEB2}) we find the Fourier transforms of the operators $
\hat{X}_{\epsilon}, \hat{P}_{\epsilon}, \hat{I}_{\epsilon}$:
\begin{eqnarray}
&& (\hat{X}_{\epsilon}\tilde{f})(k) \ = \ i \frac{d}{dk} \tilde{f}(k)\, ,
\label{MR0}
  \\
&& (\hat{P}_{\epsilon}\tilde{f})(k) \ = \ \frac{\hbar}{\epsilon}
\sin(k\epsilon)
\tilde{f}(k)\, ,  \label{MRa}\\
&& (\hat I_{\epsilon}\tilde{f})(k) \ = \ \cos(k\epsilon)
\tilde{f}(k)\, . \label{MR}
\end{eqnarray}
With the help of (\ref{MR}) we can rewrite the
commutation relation (\ref{CR}) equivalently as
\be
\left( [\hat{X}_{\epsilon}, \hat{P}_{\epsilon}] f\right)(x)
\ &=& \
 i \hbar \cos \left(\epsilon {\hat p}/{\hbar}\right) f(x)\, .
\label{20}
\ee
The latter allows to identify the lattice unit operator
$\hat{{I}}_{\epsilon}$ with $\cos \left(\epsilon {\hat
p}/{\hbar}\right)$. Indeed, $\hat{I}_{\epsilon} = \hat{\openone}$ on
all lattice nodes.

\section{Uncertainty relations on lattice\label{Sec.IV}}

We are now prepared to derive the generalized uncertainty relation
implied by the previous commutators. We shall define the uncertainty
of an observable $A$ in a state $\psi$ by the standard deviation
\begin{eqnarray}
(\Delta A)_{\psi } \ \equiv \
\sqrt{\langle \psi| (\hat{A} - \langle \psi| \hat{A}
|\psi \rangle)^2 |\psi \rangle}\, .
\end{eqnarray}
Following the conventional Robertson-Schr\"{o}dinger procedure (see, e.g.,
Ref.~\cite{Schroedinger1,Rob1,Mes1}), we derive on the spacetime lattice the inequality
\be
(\Delta X_\epsilon)_{\psi} (\Delta P_\epsilon)_{\psi} \ &\geq& \
\frac{1}{2}\left|\langle\psi|[\hat{X}_{\epsilon},
\hat{P}_{\epsilon}]|\psi\rangle\right|
\ = \ \frac{\hbar}{2}\left|\langle\psi|\hat{I}_ \epsilon |\psi\rangle\right|
\nonumber \\
&=& \ \frac{\hbar}{2}\left|\langle\psi|
\cos\left(\epsilon{\hat{p}}/{\hbar}\right)|\psi\rangle\right| \, .
\label{deltaxp}
\ee
For brevity we will omit in the following  the subscript $\psi$ in
$(\Delta A)_{\psi }$ and set $\langle \psi|\cdots |\psi \rangle
\equiv \langle \cdots \rangle_{\psi}$.

Let us now study two critical regimes of the GUP (\ref{deltaxp}):
the first is the long-wavelengths regime where $\langle  \hat p\rangle_{\psi } \rightarrow 0$;
the second regime is near the boundary of the Brillouin zone where $
\langle \hat p\rangle_{\psi}  \to \pi\hbar /2\epsilon$.
To this end we first rewrite $\langle\cos\left(\epsilon {\hat{p}}/{\hbar}\right)
\rangle_{\psi}$  as
\begin{eqnarray}
\mbox{\hspace{-5mm}}\left\langle  \cos\left(\epsilon{\hat{p}}/{\hbar}
\right)\right\rangle_{\psi}\ = \
\sum_{n =0}^{\infty} \int_{0}^{\infty} \!\!\!dp \ \! \varrho(p) \ \!
(-1)^{n}\frac{\left(\epsilon {{p}}/{\hbar}\right)^{2n}}{(2n)!}\, ,\label{II.24}
\end{eqnarray}
where $\varrho(p) \equiv |\psi(p)|^2$.

In the first case,
$\varrho(p)$ is peaked around $p \simeq 0$, so that the relation
(\ref{II.24}) becomes approximately
\be
\left\langle \cos\left(\epsilon{\hat{p}}/{\hbar}\right)\right\rangle_{\psi}
\ = \ 1 \ - \  \frac{\epsilon^2 {p}^2}{2\,\hbar^2} \ + \  {\cal{O}}{(p^4)}\, ,
\label{IV.26}
\ee
where ${p}^2 \equiv \langle  \hat{p}^2 \rangle_{\psi}$. We should
stress that expansion (\ref{IV.26}) is not an expansion in
$\epsilon$ but rather in $\epsilon p/\hbar$. So if we speak of
$\varrho(p)$ as being peaked around $p \simeq 0$ we mean that $p\ll
\hbar/\epsilon$.

Applying now  the identity
\be
\langle \hat{A}^2 \rangle_{\psi} \  = \ (\Delta {A})^2 + \langle
\hat{A}\rangle_{\psi}^2\, ,
\ee
we obtain from (\ref{deltaxp})
\be
\Delta X_\epsilon \Delta P_\epsilon \ &\gtrsim& \
\frac{\hbar}{2}\left|1 - \frac{\epsilon^2p^2}{2\hbar^2}\
\right| \nonumber \\
&=& \ \frac{\hbar}{2} \left|1 - \frac{\epsilon^2}{2\hbar^2}
\left[(\Delta p)^2 + \langle
\hat{p}\rangle^2_{\psi}\right]\right|. \ee
For mirror-symmetric states where $\langle \hat{p}\rangle_{\psi} =
0$ this implies
\be \Delta X_\epsilon \Delta P_\epsilon \ \gtrsim \
\frac{\hbar}{2}\left(1 -  \frac{\epsilon^2}{2\hbar^2}\ \!(\Delta
p)^2\right) \,. \label{II.25a} \ee
Here we have substituted $|...|$ by $(...)$ since we assume
that $\epsilon \simeq \ell_p$ (Planckian lattice)
and that $\Delta p$ is close to zero (this is our original assumption).
Therefore $\epsilon^2 (\Delta p)^2 / 2 \hbar^2 \ll 1$.

For Planckian lattices with the relation (\ref{II.26a}),
we can neglect
higher
powers of $\epsilon$
in (\ref{II.25a})
 and write
\be
\Delta X_\epsilon \Delta P_\epsilon \ \gtrsim \
\frac{\hbar}{2}\left(1 -  \frac{\epsilon^2}{2\hbar^2} \ \!(\Delta
P_\epsilon)^2\right).
\label{25}
\ee

In the second case, where $\langle \hat{p} \rangle_{\psi} \to \hbar
\pi/2\epsilon$, i.e. near the border of the Brillouin zone,  we use
the expansion:
\be
&&\mbox{\hspace{-10mm}}\langle \cos [\pi/2 + (\epsilon
\hat{p}/\hbar - \pi/2)] \rangle_\psi \ = \
\langle \sin(\pi/2 - \epsilon \hat{p}/\hbar)\rangle_\psi \nonumber \\
&&\mbox{\hspace{-0mm}} = \ \sum_{n =0}^{\infty} \int_{0}^{\infty} \!\!\!dp \ \!
\varrho(p) \ \! (-1)^{n}\frac{\left(\pi/2 - \epsilon
{p}/\hbar\right)^{2n+1}}{(2n+1)!}\, .
\label{II.29}
\ee
Under the assumption that $\varrho(p)$ is peaked
near the border of the
Brillouin zone,
the first term in the expansion
is dominant, and the uncertainty relation reduces to
\be \Delta X_\epsilon \Delta P_\epsilon \ &\geq& \
\frac{\hbar}{2}\left|\,\frac{\pi}{2} - \frac{\epsilon}{\hbar}\,
\langle\hat{p}\rangle_{\psi}\,\right|\,. \label{II.28} \ee
Since $k=p/\hbar$ lies always {\em inside\/} the Brillouin zone, we
have $\langle\hat{p}\rangle_{\psi} \leq {\pi\hbar }/{2 \epsilon }$
and can therefore  in (\ref{II.28}) substitute $|...|$ by $(...)$.
Finally, using again (\ref{II.26a}), we can write for the GUP close
to the boundary of the Brillouin zone
\be
\Delta X_\epsilon \Delta P_\epsilon \ \gtrsim \
\frac{\hbar}{2}\left(\,\frac{\pi}{2} - \frac{\epsilon}{\hbar}\,
\langle\hat{P}_\epsilon\rangle_\psi\,\right).
\label{pbig}
\ee
As the momentum reaches the boundary of the Brillouin zone, the
right-hand sides of (\ref{II.28})--(\ref{pbig}) vanish, so that
lattice quantum mechanics at short wavelengths is permitted to
exhibit classical behavior --- no irreducible lower bound for
uncertainties of two complementary observables appears!

It is worth noting that the uncertainty relation (\ref{pbig})
leads to the same physical conclusions as those found,
on a different ground, by Magueijo and
Smolin in Ref.~\cite{MS}. In particular, the world-crystal universe can become ``deterministic"
for energies near the border of the Brillouin zone, i.e.,  for
Planckian energies.

Let us remark that the scenario in which the universe at Planckian
energies is deterministic rather than being dominated by tumultuous
quantum fluctuations is a recurrent theme in 't Hooft's
``deterministic" quantum
mechanics~\cite{tHoft1,tHoft2,elze,biro,baerjee,halliwell,jizba}.

It is
straightforward
to generalize
 the above formulas
to higher dimensions.
%
In this context, a useful inequality is
\be \Delta X_\epsilon^i  \Delta |\hspace{-.8pt}{\bi {P}}_\epsilon| \ &\geq& \
\frac{\hbar}{2}| \langle\psi|({\hat{P}_\epsilon^i}/{|\hat{\bi
{P}}_\epsilon|})\cos\left(\epsilon^i{\hat{p}^i}/{\hbar}\right)
|\psi\rangle|\nonumber \\
&=& \ \frac{\hbar}{2}| \langle\psi|
\varepsilon(\hat{p}^i)\cos\left(\epsilon^i{\hat{p}^i}/{\hbar}\right)
|\psi\rangle|     \, ,\label{pbig3a} \ee
which will be needed in the following. Here
$\varepsilon(\ldots)$ is the sign function, and
\begin{eqnarray}
 |\hat{\bi
{P}}_\epsilon| \ = \ \hbar\sqrt{\sum_{j=1}^{D}
\left[\frac{\sin(\epsilon^j \hat{p}^j/\hbar)}{\epsilon^j}\right]^2
}\! .
\end{eqnarray}
Inequality
(\ref{pbig3a}) should be contrasted with inequality (\ref{deltaxp})
where the momentum is without an absolute value.

In a particular case when states $\psi$ are a combination of only
positive or only negative momentum eigenstates (e.g., incident or
reflected particle states) we can simply write
\be \Delta X_\epsilon^i \Delta |{\bi {P}}_\epsilon| \ &\geq& \
\frac{\hbar}{2}| \langle\psi|
\cos\left(\epsilon^i{\hat{p}^i}/{\hbar}\right) |\psi\rangle|
\nonumber \\[2mm]
&=& \ \frac{\hbar}{2}\left[1 - 2\left\langle
\sin^2\left(\epsilon^i{\hat{p}^i}/2{\hbar}
\right)\right\rangle_{\psi} \right] .\label{pbig4a} \ee
%

\section{Implications for Photons\label{SecV}}

We may now use the inequality
(\ref{pbig4a}) to derive the GUP for photons.

The vector potential of a photon in the Lorentz gauge in $1+1$
dimensions satisfies the wave equation
\be \frac{1}{c^2}\ \!{\partial_t^2} A^\mu (x,t) \ = \
{\partial^2_x}\,A^\mu(x,t)\, . \label{we} \ee
A plane wave solution $A^\mu(x)=
\epsilon^\mu\exp[i(kx-\omega(k)t)]$ possesses the well-known linear
dispersion relation
\be \omega(k) \ = \  c\,|k|\, , \label{LINE} \ee
with $ \epsilon ^\mu$ being a polarization vector. On a
one-dimensional lattice, the operator $\partial _x^2$ is replaced by
the lattice Laplacian $\bar\nabla\nabla$, and the spectrum becomes,
on account of Eq.~(\ref{LAST}) and (\ref{DEB2}),
\be \mbox{\hspace{-3mm}} \frac{\omega(k)}{c} =  \sqrt{K\bar K} =
\frac{\sqrt{ 2\left[ 1-\cos (k\epsilon)\right] }}{\epsilon}
=\frac{2}{\epsilon}\left|\sin \left(\frac{k \epsilon}{2}
\right)\!\right|, \ee
which reduces to (\ref{LINE}) for $\epsilon \to 0$. Denoting the
energy on the lattice $\hbar  \omega $ by $E_\epsilon$, we  obtain
the dispersion relation
\be \frac{E_\epsilon}{\hbar \,c} \ = \ \frac{2}{\epsilon}\left|\sin
\left(\frac{p\, \epsilon}{2\,\hbar} \right)\!\right|.
\label{disprel} \ee
We can also define the associated energy operator $\hat E_c$ by
replacing $p$  by $\hat p$.

For states $\psi$ with $\varrho(p)$ sharply peaked around
small $p$, we can use a spectral expansion analog of (\ref{II.24})
to obtain
\begin{eqnarray}
\Delta E_\epsilon \ \simeq \ c \Delta |p|  \ \simeq \ c \Delta|\hspace{-.8pt}P_{\epsilon}|\, . \label{epapprox1}
\end{eqnarray}
Here we have neglected higher powers of momentum and used the fact
that we deal with a  Planckian lattice. In deriving we have also
applied the cumulant expansion:
\begin{eqnarray}
\langle \hat{E}_\epsilon \rangle_{\psi} \ &=& \ (2\hbar c/\epsilon)\
\! \langle |\sin \left({\hat{p}\, \epsilon}/{2\,\hbar} \right)|
\rangle_{\psi} \nonumber \\
&=& \ \left|c\, p - c\frac{\epsilon^2 p^3}{24\, \hbar^2} +
{\cal{O}}(p^5)\right|\,. \label{epapprox}
\end{eqnarray}
%
With the help of (\ref{pbig4a}), (\ref{disprel}), and
(\ref{epapprox1}) we can  write in the long-wavelength regime
\be \Delta X_\epsilon \Delta E_\epsilon \ \geq \ \frac{\hbar
c}{2}\left[1 - \frac{\epsilon^2}{2\hbar^2 c^2}\ \! \langle
E_\epsilon^2 \rangle_{\psi}\right]. \label{26} \ee
Here $\langle E_\epsilon^2\rangle_{\psi}$ is the average quadrat of the photon energy, and
thus the square root of it can be formally identified with the energy change in the detector, i.e. $\Delta E_\epsilon$.
From this follows that if the uncertainty  of a photon position in a
state $\psi$ is $\Delta X_\epsilon$, then the energy of a detector
changes {\em at least} by amount
\be \Delta E_\epsilon \ \simeq \ \frac{\hbar c}{2}\left[1 -
\frac{\epsilon^2}{2\hbar^2 c^2}\ \! \langle E_\epsilon^2
\rangle_{\psi}\right]\!\frac{1}{\Delta X_\epsilon}\, , \label{27a}  \ee
per particle.
Remembering the Einstein relation $\Delta E = 2\pi
\hbar c/\lambda$, we can interpret $4\pi \Delta X_\epsilon$ as being
a lattice equivalent of  photon's wavelength $\lambda$.

It is interesting to observe that in the short-wavelength case we
can deduce from the exact GUP
\begin{eqnarray}
\Delta X_\epsilon \Delta |\hspace{-.8pt}{\bi {P}}_\epsilon| \ \geq \
\frac{\hbar}{2}\left[1 - \frac{\epsilon^2}{2\hbar^2 c^2}\ \! \langle
E_\epsilon^2 \rangle_{\psi} \right] ,
\end{eqnarray}
that near the border of the Brillouin zone $\Delta X_\epsilon$ takes
the approximate form (cf. Eq.~(\ref{disprel}))
\begin{eqnarray}
\Delta X_\epsilon\ &\simeq& \ \frac{\epsilon}{\pi}\left[1 \ - \
\frac{\epsilon^2}{2\hbar^2 c^2}\
\! \langle E_\epsilon^2 \rangle_{\psi} \right]\nonumber\\
&\simeq& \ \frac{\epsilon}{\pi}\left[\frac{\pi}{2} \ - \
\frac{\epsilon}{\hbar}\langle \hat{P}_{\epsilon}\rangle_{\psi}
\right].\label{V.47.a}
\end{eqnarray}
In the derivation we have used that fact that
\begin{eqnarray}
 \Delta |\hspace{-.8pt}{\bi {P}}_\epsilon| \ \leq \ \sqrt{ \langle \hat{\bi
{P}}^2_\epsilon\rangle } \ \simeq  \ \frac{\pi \hbar}{2\epsilon}\, .
\end{eqnarray}
Relation (\ref{V.47.a}) represents the smallest attainable positional
uncertainty  near the border of
the Brillouin zone. It will be
useful in the following two sections.

\section{Applications to micro black holes\label{SecVI}}

An interesting playground where one can apply the above lattice GUP's is the
hypothetical physics of micro black holes. Their mass-temperature
relation depends sensitively on the actual form of the
energy-position uncertainty relation. From this one can deduce
non-trivial phenomenological consequences. The passage from the
energy-position uncertainty relation to the micro black hole
mass-temperature relation has been intensively studied in recent
years. For definiteness we shall follow here
 the
treatment of
Refs.~\cite{FS9506,ACSantiago,Adler2,CavagliaD,CDM03,Susskind,nouicer,Glimpses}.
An alternative derivation based on the so-called  Landauer principle
will be
presented in Appendix.

We start with an assumption that the lattice spacing is roughly of
order Planck length, i.e., $\epsilon = a \ell_p$, where $a>0$ is of
order of unity. Let us now imagine that we have found a black hole
on the lattice as a discretized version of a Schwarzschild solution.
It is a pile up of disclinations. If the Schwarzschild radius is
much larger than the lattice spacing $\epsilon $, this will not look
much different from the well-known continuum solution. We must avoid
too small black holes, for otherwise, completely new physics will
set in near the center, due to the high  concentration of defects.
These will cause the ``melting" of the world crystal at a critical
defect density~\cite{MELT}, and the emerging trans-horizon general
relativity would look completely different from Einstein's theory.

Following the classical argument of the Heisenberg
microscope~\cite{Heisenberg}, we know that the smallest resolvable
detail $\delta x$ of an object goes roughly as the wavelength of the
employed photons. If $E$ is the (average) energy of the photons used
in the microscope, then
\be \delta x \ \simeq \ \frac{\hbar c}{2 E}\,. \label{He} \ee
Conversely, with the relation (\ref{He}) one can compute the energy
$E$ of a photon with a given (average) wavelength $\lambda \simeq
\delta x$. As a consequence of Eq.~(\ref{26}), we can write the
lattice version of this standard Heisenberg formula as
\be \delta X_\epsilon \ \simeq \ \frac{\hbar c}{2 E_\epsilon}
\left[1 - \frac{\epsilon^2}{2\hbar^2 c^2}\ \!(E_\epsilon)^2
\right]\! , \label{43} \ee
which links the (average) wavelength of a photon to its energy
$E_\epsilon$. Since the lattice spacing is $\epsilon = a \ell_p$ and
the Planck energy ${\cal E}_p \ = \ \hbar c / 2\ell_p$,
Eq.~(\ref{43}) can be rewritten as
\be \delta X_\epsilon \ \simeq \ \frac{\hbar c}{2 E_\epsilon}
 - \frac{a^2\ell_p E_\epsilon}{8 {\cal E}_p}\, .
\label{45} \ee
Let us now loosely follow the argument of
Refs.~\cite{FS9506,ACSantiago,Adler2,CavagliaD,CDM03,Susskind,nouicer,Glimpses}
and consider an ensemble of unpolarized photons of Hawking radiation
just outside the event horizon. From a geometrical point of view,
it's easy to see that the position uncertainty of such photons is of
the order of the Schwarzschild radius $R_S$ of the hole. An
equivalent argument comes from considering the average wavelength of
the Hawking radiation, which is of the order of the geometrical size
of the hole (see e.g. Ref.~\cite{Susskind}, chapter 5). By recalling
that $R_S=\ell_p m$, where $m=M/M_p$ is the black hole mass in
Planck units ($M_p={\cal E}_p/c^2$), we can estimate the photon
positional uncertainty as
\be \delta X_\epsilon \ \simeq \ 2\mu R_S \ =  \ 2\mu \ell_p m\, .
\label{VI.48a} \ee
The proportionality constant  $\mu$ is of order unity and will be
fixed shortly. According to the above arguments, $m$ must be assumed
to be much larger than unity, in order to avoid the melting
transition. With (\ref{VI.48a}) we can rephrase Eq.~(\ref{45}) as
\be 2\mu m \ \simeq \ \frac{{\cal E}_p}{ E_\epsilon} - \frac{a^2}{8}
\frac{E_\epsilon}{{\cal E}_p}\, . \label{47} \ee
According to the equipartition principle the average energy
$E_\epsilon$ of unpolarized photons of the Hawking radiation is
linked with their temperature $T$ as
\be E_\epsilon \ = \ k_B T\, . \ee
In order to fix $\mu$, we go to the continuum lattice limit
$\epsilon \to 0$ ($a \to 0$), and require that
 formula (\ref{47}) predicts the standard semiclassical
Hawking temperature:
%
\be T_{H} \ = \ \frac{\hbar c^3}{8\pi G k_B M} \ = \ \frac{\hbar
c}{4 \pi k_B R_S}\, . \label{Hw} \ee
This fixes $\mu = \pi$.

Defining the Planck temperature $T_p$ so that ${\cal E}_p = k_B
T_p/2$ and measuring all temperatures in Planck units as $\Theta =
T/T_p$, we can finally cast formula (\ref{47}) in the form
\be 2m \ = \ \frac{1}{2\pi\Theta} \,-\, \zeta^2\, 2 \pi \Theta\, ,
\label{MT-} \ee
where we have defined the {\em deformation} parameter $\zeta =
a/(2\pi\sqrt{2})$.

As already mentioned, in the continuum limit both $\epsilon$ and $a$
tend to zero and (\ref{45}) reduces to the ordinary Heisenberg
uncertainty principle. In this case Eq.~(\ref{MT-}) boils down to
\be m \ =  \ \frac{1}{4 \pi \Theta}\, . \label{H} \ee
This is the dimensionless version of Hawking's formula (\ref{Hw})
for {\em large} black holes.

Historically, the validity of (\ref{Hw}) was also postulated for
micro black holes on the assumption that the black hole
thermodynamics is universally valid for any black hole, be it formed
via star collapse, or  primordially via quantum fluctuations. Such
an assumption is by no means warranted without some further input
about mesoscopic and/or microscopic energy scales (much like in
ordinary thermodynamics) and, in fact, we have seen that corrections
should be expected at short world-crystal scales.

It is instructive to compare our mass-temperature relation
(\ref{MT-}) with the one suggested by the so-called stringy
uncertainty relation~\cite{veneziano,mascad}. There the sign of the
correction term in   (\ref{MT-}) is positive:
\be 2 m \ = \ \frac{1}{2 \pi \Theta} + \zeta^2 \,2 \pi \Theta\,.
\label{MT+} \ee

\begin{figure}[h]
\centerline{\epsfxsize=2.9truein\epsfysize=1.8truein\epsfbox{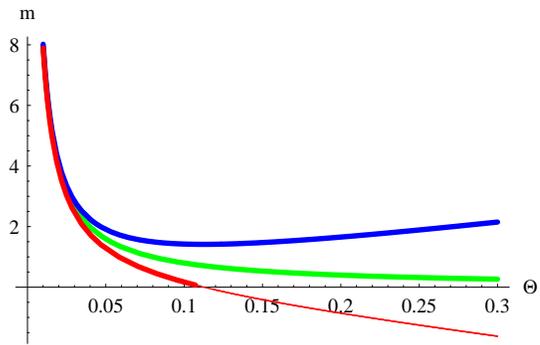}}
\caption[]{Diagrams for the three {\em mass-temperature} relations,
ours (red), Hawking's (green), and stringy GUP result (blue), with
$\zeta=\sqrt{2}$, as an example. As a consequence of
the lattice
uncertainty principle the evaporation ends at a {\em finite\/} temperature
with a zero rest-mass remnant.} \vspace{0.2cm} \hrule
\label{MTfig}
\end{figure}

The phenomenological consequences of the relation (\ref{MT-}) are
quite different from those of the stringy result (\ref{MT+}). In
Fig.~\ref{MTfig} we compare the two results, and add also the curve
for the ordinary Hawking relation (\ref{H}).
Considering $m$ and $\Theta$ as functions of time, we can follow the
evolution of a micro black hole from the curves in Fig.~1. For the
stringy GUP, the blue line predicts a maximum temperature
\begin{subequations}
\be \Theta_{{\rm max}} \ = \ \frac{1}{2 \pi {\zeta}}\, , \ee
and a minimum rest mass
\be m_{{\rm min}}\ = \ {\zeta}\,. \ee \label{Thmax}
\end{subequations}
The end of the evaporation process is reached in a finite time, the
final temperature is finite, and there is a remnant of a finite rest
mass (cf.
Refs.~\cite{FS9506,ACSantiago,Adler2,CavagliaD,CDM03,nouicer,Glimpses,Casadio}).

From the standard Heisenberg uncertainty principle we find the green
curve, representing the usual Hawking formula. Here the evaporation
process ends, after a finite time, with a zero mass and a worrisome
infinite temperature. In the literature, the undesired {\em
infinite} final temperature predicted by Hawking's formula has so
far been cured only with the help of the stringy GUP, which brings
the final temperature to a finite value. This result is, however,
also questionable since it implies the existence of finite-mass
remnants in the universe. Though by some authors such remnants are
greeted as relevant candidates for dark matter~\cite{chenDM}, others
point out that their existence would create further complications
such as the entropy/information problem~\cite{beken}, detectability
issue, or their (excessive) production in the early universe
~\cite{Giddings:08,remnants}.

In contrast to these results, our lattice GUP predicts the red
curve. This yields a finite end temperature
\be \Theta_{{\rm max}}\ = \ \frac{1}{2 \pi {\zeta}}\, , \ee
with a zero-mass remnant. The mass-temperature formula  (\ref{MT-})
thus solves at once several
 problems by predicting the end of the
evaporation process at a {\em finite} final temperature with {\em
zero}-mass remnants.

Is should be stressed that since the photon GUP (\ref{26}) and
(\ref{43}) holds only for states $\psi$ where $\langle \hat{p}^2
\rangle_{\psi} \ll \hbar^2/\epsilon^2$, our reasonings are warranted
only for
\begin{eqnarray}
E_\epsilon \ \ll \  \hbar c/\epsilon \ \simeq \ \mathcal{E}_p/a
\;\;\;\; \Rightarrow \;\;\;\; 2\pi \zeta \ \ll \  \frac{1}{\Theta}
\, .
\end{eqnarray}
This implies, in particular,  that when $\Theta$ is close to
$\Theta_{{\rm max}}$ our long-wavelength approximation cannot be
trusted.

To understand the behavior of the system close to $\Theta_{{\rm
max}}$ we must turn to the short-wavelength limit, Eqs.~(\ref{pbig})
and (\ref{V.47.a}). In this regime the  momenta lie close to the
border of the Brillouin zone $ \langle \hat p\rangle_\psi \simeq
\langle\hat{P}_{\epsilon}\rangle_\psi \simeq \pi \hbar/(2\epsilon)$,
and Eq.~(\ref{disprel}) implies
\be E_\epsilon\  \simeq \ \frac{\sqrt{2}}{\epsilon}\,\hbar c \, ,
\ee
which for the Planckian lattice, where $\epsilon = a \,\ell_p$,
gives $E_\epsilon \simeq \mathcal{E}_p/(\pi\, \zeta)$. Considering
again the uncertainty in the photon position as $\delta X_\epsilon
\simeq 2 \pi \ell_p m$ (cf. Eq.~(\ref{VI.48a})), GUP (\ref{V.47.a})
then predicts
\be \delta X_\epsilon \ \propto \ m \ \simeq \ \frac{a}{2\pi^2}
\left(1 - \frac{a^2 \ell_p^2}{2\hbar^2c^2}\frac{{\cal
E}_p^2}{\pi^2\zeta^2}\,\right) \ =\ 0\, . \ee
%
%
We can thus conclude that the mass of the micro black hole must go
to zero. This is also consistent with our previous long-wavelength
considerations. The micro black hole therefore evaporates
completely, without leaving remnants.



%
%
%
%
\section{Entropy and heat capacity\label{entr}}
%
%
In this section we exhibit
the modified thermodynamic entropy and heat
capacity of a black hole implied by
the new mass-temperature formula
(\ref{MT-}).
%
%
\subsection{Entropy}
%
%
From the first law of black hole thermodynamics~\cite{BCH} we know
that the differential of the thermodynamical entropy of a
Schwarzschild black hole reads
\be dS \ =  \ \frac{dE}{T_H}\, , \label{bht} \ee
where $dE$ is the amount of energy swallowed by a black hole with
Hawking temperature $T_H$. In Eq. (\ref{bht}) the increase in the
internal energy is equal to the added heat because
a black hole makes no mechanical work when its entropy/surface
changes (expanding surface does not exert any pressure).

Rewriting Eq.~(\ref{bht}) with the dimensionless variables $m$ and
$\Theta$ we get
\be d S \ = \ \frac{k_B}{2} \frac{d m}{\Theta} \, . \ee
Inserting here formula (\ref{MT-}) we find
\be d S \ = \ \frac{k_B}{2} \frac{d m}{\Theta} \ = \ -
\frac{k_B}{4}\left(\frac{1}{2\pi\Theta^3} +
\frac{2\pi\zeta^2}{\Theta}\right)\,d\Theta \,. \label{ds} \ee
By integrating $dS$ we obtain $S=S(\Theta)$. Just as formula
(\ref{MT-}), the relation (\ref{ds}) can be trusted only for $\Theta
\ll \Theta_{\rm max} = 1/2\pi\zeta$. Thus, when integrating
(\ref{ds}), we should do this only up to a cutoff
$\tilde{\Theta}_{\rm max} \ll \Theta_{\rm max}$. The additive
constant in $S$ can be then be fixed by requiring that $S=0$ when
$\Theta \to \tilde{\Theta}_{\rm max}$. This is equivalent to what is
usually done when calculating the Hawking temperature for a
Schwarzschild black hole. There one fixes the additive constant in
the entropy integral to be zero for $m=0$, so that $S(m=0) =
S(\Theta \rightarrow \infty) =0$ (the minimum mass attainable in the
standard Hawking effect is $m=0$). Thus we obtain
\be S\ = \ \frac{k_B}{4} \int_{\Theta}^{\tilde{\Theta}_{\rm max}}
\left(\frac{1}{2\pi\Theta'^3} +
\frac{2\pi\zeta^2}{\Theta'}\right)\,d\Theta' \label{Sint} \ee
where the sign was chosen in order to have a positive entropy.
%

The integral (\ref{Sint}) yields
\be S(\Theta)\ = \ \frac{k_B}{16\, \pi}\left(\frac{1}{\Theta^2} -
\frac{1}{\tilde{\Theta}_{\rm max}^2} + 8\pi^2\zeta^2\log
\frac{\tilde{\Theta}_{\rm max}}{\Theta}\right)\! . \label{S4} \ee
The entropy is always positive,
and $S \to 0$ for $\Theta \to
\tilde{\Theta}_{\rm max}$.

%
%
%
%
\subsection{Heat Capacity}
%
%
With entropy formulae (\ref{ds}) and (\ref{S4}) at hand we can now
compute the heat capacity of a (micro) black hole in the
world-crystal. This will give us important insights on the final
stage of the evaporation process. Again, we shall obtain formulae
valid only for $\Theta \ll \Theta_{\rm max} = 1/(2\pi\zeta)$.

The heat capacity $C$ of a black hole is defined via the relation
\be dQ \ = \ dE \ = \ C dT\, . \label{VII.B.1} \ee
%
The pressure exerted on the environment by the expanding black hole
surface is zero. Hence we do not need to specify which $C$ is meant.
%
%

With the help of (\ref{bht}) and (\ref{VII.B.1}) we obtain
\begin{eqnarray}
C \ = \ T \left(\frac{d S}{ d T}\right) \ = \ \Theta \left(\frac{d
S}{ d \Theta}\right)\, ,
\end{eqnarray}
%
which yields
\be C \ = \  -\frac{\pi k_B}{2}\left[\zeta^2 +
\frac{1}{(2\pi\Theta)^2}\right]\! . \label{VII.B.4} \ee
From this clearly follows that $C$ is always negative.

Most condensed-matter systems have $C>0$. However, because of
instabilities induced by gravity this is generally not the case in
astrophysics~\cite{thirring,zeh}, especially in black hole physics.
A Schwarzschild black hole has $C<0$ which indicates that the black
hole becomes hotter by radiating. The result (\ref{VII.B.4}) implies
that this scenario holds also for micro black holes in the
world-crystal.

%
In case of stringy GUP, we have to use Eq. (\ref{MT+}) as the
mass-temperature formula. The expression for the heat capacity then
reads
\be C \ = \ \frac{\pi k_B}{2}\left[\zeta^2 -
\frac{1}{(2\pi\Theta)^2}\right]\! . \label{HC} \ee
Since also here $0<\Theta<\Theta_{\rm max}= 1/2\pi\zeta$, black
holes have negative specific heat also according to the stringy GUP.
However, stringy GUP displays a striking difference with respect to
lattice GUP. In fact, since in principle we can trust Eq. (\ref{MT+})
also when $\Theta \simeq \Theta_{\rm max}=1/2\pi\zeta$, then from
(\ref{HC}) we have, in such limit, $C=0$. This means that for the
stringy GUP the specific heat vanishes at the end point of the
evaporation process in a finite time, so that the black hole at the
end of its evolution cannot exchange energy with the surrounding
space. In other words, the black hole stops to interact
thermodynamically with the environment. The final stage of the
Hawking evaporation, according to the stringy GUP scenario, contains
a Planck-size remnant with a maximal temperature $\Theta=\Theta_{\rm
max}$, but thermodynamically inert. The remnant behaves like an
elementary particle --- there are no internal degrees of freedom to
excite in order to produce a heat absorption or emission.

To understand the heat exchange in the last live stage of the
world-crystal black hole we cannot use the long-wavelength formula
(\ref{VII.B.4}). Instead, we must turn to Eq.~(\ref{V.47.a}). Since
in our scenario the micro black hole disappears at the critical
temperature $\Theta_{\rm max}$, it is more appropriate to write the
mass-temperature formula (\ref{V.47.a}) in the form
\begin{eqnarray}
m \ \simeq \ \theta(\Theta_{\rm max} - \Theta)\ \!
\frac{\sqrt{2}}{\pi}\zeta \left(1 \ - \ \frac{a^2}{8}\ \!
\Theta^2\right)\! ,
\end{eqnarray}
where $\theta(t)$ is the Heaviside step function. For the
specific heat this implies
\begin{eqnarray}
C \ \simeq \ -  \theta(\Theta_{\rm max} - \Theta)\ \! \sqrt{2} \pi \
\!k_B \zeta^3 \ \! \Theta\, . \label{VII.B.69}
\end{eqnarray}
So, in contrast to the stringy result, a world-crystal black hole
exchanges heat with its environment by radiation until the last moment
of its existence, and unlike the Schwarzschild black hole, the heat
exchange with the environment increases in the final stage of its
evaporation. In addition, because of the $\theta$-function in
(\ref{VII.B.69}), the transition from the universe with the
world-crystal black hole to the one without it is of first order.

\section{Further applications\label{appl}}

So far we have studied the consequence of the GUP on the micro black
holes. Let us briefly mention two further applications.

\subsection{'t~Hooft's proposal\label{dics1}}

The first application relates to 't~Hooft's proposal which purports
to justify that our quantum world is merely a low-energy limit of a
deterministic system operating at a deeper, perhaps Planckian, level
of dynamics~\cite{tHoft1,tHoft2}. As a deterministic substrate
't~Hooft has proposed various cellular automata (CA) models.

In general, a CA is an array of cells forming a discrete lattice. All
cells are typically equivalent and can take one of a finite number
of possible {\em discrete states}. Like space, time is discrete as
well. At each time step every cell updates its state according to a
{\em transition rule} which takes into account the previous states
of cells in the neighborhood, including its own state. In this sense
the evolution is deterministic.

\begin{figure}[h]
\centerline{\epsfysize=1.3truein\epsfbox{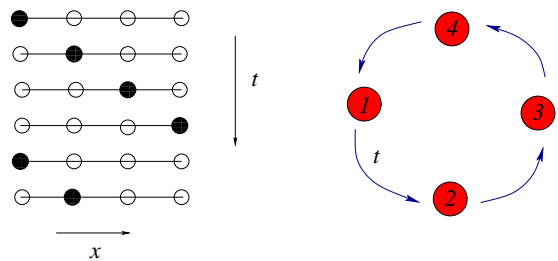}}
\caption[]{CA with discrete time evolution described by Eq.~
(\ref{disc73}) and with the periodicity condition $\sigma_i =
\sigma_{i+4}$. The right-hand shows an equivalent graphical
representation.} \vspace{0.2cm} \hrule
\label{VII.1}
\end{figure}

One of the simplest CA considered by 't~Hooft is the $1$-dimensional
periodic CA with $4$-state cells,
 and with the nearest neighbor (N-N)
transition rule, see Fig.~\ref{VII.1}.
This ``clock like" CA can be generally described with $2N+1$ cells
$\sigma_i ~(i = -N, \ldots, N)$ each with two possible states
$\{0,1\}$ (cell is, e.g.,  white or black). The discrete time
evolution with the elementary time step $\delta t$ is described by
the N-N transition rule
\begin{eqnarray}
&&\mbox{\hspace{-5mm}}(\sigma_{i-1}, \sigma_i, \sigma_{i+1}) \
\rightarrow \ \sigma'_{i}
\ = \ \sigma_i(t + \delta t):\nonumber \\[2mm]
&&\mbox{\hspace{-5mm}}(0,0,0) \rightarrow 0, \;\; (0,0,1)
\rightarrow 0, \;\; (0,1,0) \rightarrow 0,\nonumber \\[1mm]
&&\mbox{\hspace{-5mm}}(1,0,0) \rightarrow 1, \;\;(1,1,0) \rightarrow
0, \;\; (0,1,1) \rightarrow 0,\nonumber \\[1mm]
&&\mbox{\hspace{-5mm}}(1,0,1) \rightarrow 0,\;\; (1,1,1) \rightarrow
0\, ,\label{disc73}
\end{eqnarray}
with the Born-von Karman periodicity condition $\sigma_i =
\sigma_{i+2N+1}$.

The cells can be algebraically represented by orthonormal vectors
\begin{eqnarray*}
\sigma_{-N}  =  \left(\ba{c} 0 \\ 0\\ \vdots \\
1\ea\right)\!
;\;\sigma_{-N+1}  =  \left(\ba{c} 1 \\ 0\\ \vdots \\
0\ea\right)\!;\;\ldots \; \sigma_{N}=\left(\ba{c} 0 \\ \vdots \\ 1\\
0\ea\right)\! ,
\end{eqnarray*}
$\sigma_{-N} = \sigma_{N+1}$. On the basis spanned by  $\sigma_i$
the elementary-time step evolution operator is:
\begin{eqnarray}
\mbox{\hspace{-2mm}} \hat{U}(\delta t =\tau) = e^{-i \hat{H} \delta
t} = e^{-i\frac{\pi}{2N+1}}\! \lf(\ba{ccccc} 0 &&&&1 \\ 1&0&&&
\\ &&\ddots&\ddots& \\&&&1&0 \ea \ri)\! , \label{VIII.74.a}
\end{eqnarray}
which, among others, defines the Hamiltonian $\hat{H}$. The
pre-factor $e^{-i\frac{\pi}{2N+1}}$  is known as 't~Hooft's phase
convention. Because $\hat{U}^{{2N+1}} = - \hat{\openone}$  one can
diagonalize $\hat{U}$ as
\begin{eqnarray*}
\hat{U}_{\rm{diag}} \ = \ e^{-i\frac{\pi}{2N+1}}\
\mbox{diag}\!\left(e^{-i\frac{2\pi N}{2N+1}},\ldots,1,\ldots,
e^{i\frac{2 \pi N}{2N+1}}\right)\, .
\end{eqnarray*}
If we denote the eigenstates of $\hat{H}$ as $|n\ran$, we find that
\begin{eqnarray}
|n\rangle \ = \  \frac{1}{\sqrt{2N+1}}\sum_{\ell=-N}^{N} \exp\left[-{i
\frac{2\pi n}{2N+1} \ \!\ell}\right] \sigma_{\ell}\, ,
\end{eqnarray}
with $n = -N,\ldots ,N$ and the ensuing ``energy'' spectrum reads
\begin{eqnarray}
\hat{H} |n\rangle \ = \ \om_{N}\, \left(n+
\mbox{$\frac{1}{2}$}\right) |n\rangle , \;\;\;\; \omega_{N} \equiv
\frac{2\pi}{(2N+1) \delta t}\, . \label{VII.76.b}
\end{eqnarray}
The energy values $E_n$ resemble the spectrum of the harmonic
oscillator, except that the $n$'s are bounded and can attain
negative values. We shall be coming back to this issue shortly.

Position and momentum can be represented by operators with the
matrix representations
\begin{eqnarray}
\hat{X}_{\epsilon} &=& \lf(\ba{ccccc} (-N+1)\epsilon&0&0&\cdots&0 \\
0&(-N +2)\epsilon&0&&0
\\ \vdots&&\ddots&&\vdots \\0&\cdots&&N\epsilon&0\\0&\cdots&&0&-N\epsilon \ea \ri)\! ,
\nonumber \\[2mm]
\hat{P}_{\epsilon} &=& \ \frac{1}{2i\epsilon}\!\lf(\ba{cccccc}
0&1&0&0&\cdots&-1 \\ -1&0&1&0&\cdots&0 \\
0&-1&0&1&\cdots&0 \\ \vdots&&&\ddots&&\vdots  \\
0&0&\cdots&&0&1\\1&0&\cdots&&-1&0 \ea \ri)\! ,\label{VIII.77.a}
\end{eqnarray}
where $\epsilon = 2\pi/(2N+1)$. With these we obtain the commutator
\begin{eqnarray}
[\hat{X}_{\epsilon},\hat{P}_{\epsilon}]  \ = \ \frac{i}{2} \lf(\ba{cccccc}
0&1&0&0&\cdots&1 \\ 1&0&1&0&\cdots&0 \\
0&1&0&1&\cdots&0 \\ \vdots&&&\ddots&&\vdots
\\ 0&0&\cdots&&0&1\\1&0&\cdots&&1&0 \ea \ri)\! . \label{VIII.78.a}
\end{eqnarray}
In deriving (\ref{VIII.78.a}) we have used the the periodicity
condition $\sigma_{-1} = \sigma_{2N}$. By comparing this with the
evolution operator (\ref{VIII.74.a}) we have
\begin{eqnarray}
[\hat{X}_{\epsilon},\hat{P}_{\epsilon}]  \ = \    i\cos[(\hat{H} -
\omega_N/2) \delta t]\, ,
\end{eqnarray}
which for small $\delta t$ gives
\begin{eqnarray}
[\hat{X}_{\epsilon},\hat{P}_{\epsilon}]  \ \simeq \  i \left[1-
\frac{\delta t^2}{2} (\hat{H} - \omega_N/2)^2 \right]\! .
\end{eqnarray}
In addition,
we deduce
from (\ref{VIII.74.a}) and (\ref{VIII.77.a})
 that
\begin{eqnarray}
\sin[(\hat{H} - \omega_N/2) \delta t] \ = \ \epsilon
\hat{P}_{\epsilon}\, , \label{VIII.81.a}
\end{eqnarray}
which is compatible with the result (\ref{MRa}). From
(\ref{VIII.81.a}) follows that $\hat{H}$ depends only on
$\hat{P}_{\epsilon}$ but not on $\hat{X}_{\epsilon}$. So
$\hat{P}_{\epsilon}$ and $\hat{H}$ are simultaneously
diagonalizable. By defining the operators $\hat{K}_{\pm}$ as
\begin{eqnarray}
\hat{K}_{+} \ &=& \ e^{-i \hat{X}_{\epsilon}} \hat{P}_{\epsilon}, \;\;\;\;\;
\hat{K}_{-} \ = \ \hat{P}_{\epsilon}\ \! e^{ i \hat{X}_{\epsilon}}\, ,
\end{eqnarray}
($\hat{K}_{-} = \hat{K}_{+}^{\dag}$), so that
\begin{eqnarray}
\mbox{\hspace{-6mm}}\hat{K}_{+}|n \rangle &=& \frac{2N +1}{2\pi} \sin\left[\frac{2\pi}{2N +1}\ \! n \right]
\ \!|n+1 \rangle\, , \nonumber \\
\mbox{\hspace{-6mm}}\hat{K}_{-}|n \rangle &=&  \frac{2N +1}{2\pi} \sin\left[\frac{2\pi}{2N +1} (n-1) \right]
\ \!|n-1 \rangle\, ,
\label{VIII.84.bb}
\end{eqnarray}
one can persuade itself that $\hat{H}$ and $\hat{K}_{\pm}$ close the
deformed algebra
%
%
which in the large $N$ limit (i.e., in the small $\epsilon$ or $\delta t$ limit) reduces to
\begin{eqnarray}
[\hat{H}, \hat{K}_{\pm}] \ = \ \pm \ \!\omega \hat{K}_{\pm}, \;\;\;\;\; [\hat{K}_{+},\hat{K}_{-}] \ = \ -\frac{2
\hat{H}}{\,\omega}  \, ,
\label{VIII.84.b}
\end{eqnarray}
with $\omega = \omega_{\infty}$.

Note that for large $N$ one can identify
(\ref{VII.76.b}), (\ref{VIII.84.bb}), and (\ref{VIII.84.b}) with the
representation of $SU(1,1)$ known as the discrete
series $D^+_{1/2}\oplus D^-_{1/2}$ (cf. Ref.~\cite{JBV}). Generally,
the Lie algebra $D^+_k\oplus D^-_k$ is defined through the relations:
\begin{eqnarray}
&&\hat{L}_3 |k,m\rangle  \ = \ (m+k)|k,m\rangle \, ,
\nonumber \\[1mm]  &&\hat{L}_+ |k,m\rangle \
= \ \sqrt{(m+2k)(m+1)}\ |k,m+1\rangle \, , \nonumber \\[1mm]
&&\hat{L}_- |k,m\rangle \ = \ \sqrt{(n+2k-1)m}\ |k,m-1\rangle \, ,
\nonumber \\[1mm]
&&[\hat{L}_3, \hat{L}_\pm] \ = \ \pm \hat{L}_\pm , \;\;\;\;\;
[\hat{L}_+,\hat{L}_-]\ = \ - 2\hat{L}_3\, .
\end{eqnarray}
Here $m+k  = \pm k, \pm (k+1), \pm (k+2), \ldots$ and $k= \frac{1}{2},1,\frac{3}{2},2, \ldots$ is
the so-called Bargmann index which labels the representations.
From this we have that $D^+_{1/2}\oplus D^-_{1/2}$ corresponds to
\begin{eqnarray}
&&\hat{L}_3 |\sfrac{1}{2},m\rangle \ = \ (m+1/2)|\sfrac{1}{2},m\rangle\,,\non \\
&&\hat{L}_+ |\sfrac{1}{2},m\rangle \ = \ (m+1) |\sfrac{1}{2},m+1\rangle\,, \non \\
&&\hat{L}_- |\sfrac{1}{2},m\rangle \ = \ m |\sfrac{1}{2},m-1\rangle\, , \label{ho}
\end{eqnarray}
Identification with the large-$N$ limit  of (\ref{VII.76.b}) and
(\ref{VIII.84.bb}) is established when we identify $\hat{H}$ in
(\ref{ho}) with $\omega \hat{L}_{3}$, $\hat{K}_{\pm}$ with
$\hat{L}_{\pm}$, and set $m =n$.


At this stage one can invoke 't~Hooft's loss of information
condition~\cite{tHoft1,tHoft2}, and project out the negative part of
the spectra. A plausible
rationale for this step can be found, e.g., in irreversibility of
computational process due to a finite  storage capacity~\cite{JBV2}.

After the negative energy spectrum
is removed (erased), we obtain only the positive discrete series
$D_{{1}/{2}}^+$ (i.e., representation where $m= 0,1,2,\ldots$), and the
Hamiltonian morphs into a non-negative spectrum Hamiltonian $H^+$.

The usual $W(1)$-algebra  of the quantum harmonic oscillator emerges
after we introduce the following mapping in the universal
enveloping algebra of $SU(1,1)$:
\begin{eqnarray} \lab{holstein}
\mbox{\hspace{-4mm}}\hat{a} \ = \ \frac{1}{\sqrt{\hat{L}_3 + 1/2}} \, \hat{L}_-, \;\;\;\;
\hat{a}^\dag \ = \ \hat{L}_+ \, \frac{1}{\sqrt{\hat{L}_3 + 1/2}}\, .
\end{eqnarray}
The latter gives a one-to-one (non-linear) mapping between the
deterministic cellular automaton system (with information loss)
and the quantum harmonic oscillator. The reader will recognize the mapping
Eq. (\ref{holstein}) as the non-compact analog~\cite{gerry} of the
well-known Holstein-Primakoff representation for $SU(2)$ spin
systems~\cite{holstein}.

Our
operators $\hat{X}_{\epsilon}$, $\hat{P}_{\epsilon}$
and $\hat{I}_{\epsilon}$ may be viewed
as
the same
cellular automaton
$E(2)$ algebra as discussed in Section~\ref{Sec.III}.
Thus
we can conclude that in the Planckian
scale the system must behave deterministically --- which
is one of the defining property of
cellular automata. It is only at low energies when the loss of information leads to the emergent
degrees of freedom resulting
in the usual quantum mechanical description

\subsection{Double special relativity\label{dics1a}}

The second application relates to the idea of double (or doubly or
deformed) special relativity (DSR) (see, e.g., Refs.~\cite{DSR,MS}).
The general idea is that if the Planck length is a truly  universal
quantity, then it should look the same to any inertial observer.
This demands a modification (deformation) of the Lorenz
transformations, to accommodate an invariant length scale. In
Ref.~\cite{MS} the nonlinearity of the deformed Lorenz
transformations lead the authors to novel commutators between
spacetime coordinates and momenta, depending on the energy
\be
\left[{\hat{x}}^i, {\hat{p}}_j\right]
= {\rm i} \hbar \left(1-\frac{E}{{\cal E}_p}\right) \delta^i_j\, ,
\label{defcomm}
\ee
where $E$ is the energy scale of the particle to which the deformed
Lorenz boost is to be applied, while ${\cal E}_p$ is the Planck
energy. This suggests that they have an energy-dependent Planck
``constant" $\hbar(E)=\hbar (1-E/{\cal E}_p)$. Their model also
implies that $\hbar(E) \to 0$ for $E \to {\cal E}_p$. For energies
much below that Planck regime, the usual Heisenberg commutators are
recovered, but when $E \simeq {\cal E}_p$ one has $\hbar({\cal E}_p)
\simeq 0$. So the Planck energy is not only an invariant in this
model, but the world looks also apparently classical at the Planck
scale, similarly as in 't~Hooft's proposal.

The connections of the DSR model with our proposal are at this point self evident.
Our GUP (\ref{20}), (\ref{deltaxp}) implies that,
at the boundary of the Brillouin zone, when $\langle \hat{p}
\rangle_{\psi} \to \hbar \pi/2\epsilon$, i.e. for Planck energies
$E_\epsilon \simeq (2\sqrt{2}/a){\cal E}_p$,
the fundamental commutator vanishes
\be
 [\hat{X}_{\epsilon}, \hat{P}_{\epsilon}] \simeq 0\, ,
\ee and since \be \Delta X_\epsilon \Delta P_\epsilon \ \gtrsim \
0\, , \ee
lattice quantum mechanics at short wavelengths allows for classical
behavior, that is uncertainties of two complementary observables
{\em can} be {\em simultaneously} zero.

However, if we express the fundamental commutator (\ref{20}) of our model in terms of energy,
using the exact relation (\ref{disprel}), we find (for $\epsilon = a \ell_p$)
\be
 \left[\hat{X}_{\epsilon}, \hat{P}_{\epsilon}\right] f(x)
\ &=& \
 i \hbar \left(1 - \frac{a^2}{8}\frac{\hat{E}^2}{{\cal E}^2_p}\right) f(x)\, .
\ee
This means that the deforming term in our model is quadratic in the energy,
instead of
the linear dependence in the energy of the DSR model (\ref{defcomm}).


\section{Discussion and Summary\label{dics2a}}

It should be noted that the present lattice generalization of the
uncertainty principle is not an approximate description, but it is
an exact formula necessarily implied by our model of lattice space
time. The great majority of the GUP research has always borrowed the
deformed commutator $[\hat{x},\hat{p}] = i\hbar(1+\kappa \hat{p}^2)$
either from string theory, or from heuristic arguments about black
holes~\cite{veneziano,mascad}. To be precise, even in string
theory~\cite{veneziano} the formula expressing the GUP is not
derived from the basic features of the model, but instead it is
deduced from high-energy gedanken experiments of string scatterings.
In contrast to this we have {\em derived\/} all results from a
simple lattice model of spacetime, and from the analytic structure of the
basic commutator (\ref{20}).

We have calculated the uncertainties on a crystal-like universe
whose lattice spacing is of the order of Planck length --- the
so-called world crystal. When the energies lie near the border of
the Brillouin zone, i.e., for Planckian energies, the uncertainty
relations for position and momenta do not pose any lower bound on
the associated uncertainties. Hence the world crystal universe can
become``deterministic'' at Planckian energies. In this high-energy
regime, our lattice uncertainty relations resemble the double
special relativity result of Magueijo and Smolin.

The scenario in which the universe
at Planckian energies is deterministic rather than being dominated
by quantum fluctuations is a starting point in 't~Hooft's
``deterministic" quantum mechanics.

With the generalized uncertainty relation at hand we have been able
to derive a new mass-temperature relation for Schwarzschild micro
black holes. In contrast to standard results based on Heisenberg or
stringy uncertainty relations, our mass-temperature formula predicts
both finite Hawking's temperature and a zero rest-mass remnant at
the end of the evaporation process. Especially the absence of
remnants is a welcome bonus which allows to avoid such conceptual
difficulties as entropy/information problem or why we do not
experimentally observe the remnants that must have been prodigiously
produced in the early universe.

Apart from the mass-temperature relation we have also computed two
relevant thermodynamic characteristics, namely entropy and heat
capacity. Particularly the heat capacity provided an important
insight into the last life stage of the world-crystal micro black
holes. In contrast to the stringy result, our result indicates that
world-crystal micro black hole exchanges heat with its  environment
(radiate) till the last moment of its existence, and unlike the
Schwarzschild micro black hole, the heat exchange with the
environment increases in the final stage of the evaporation. In
addition, the transition from the universe with the world-crystal
micro black hole to the one without is of the first order.

Since
the world crystal
physics allows for deterministic description
 of the physics at Planckian energies,
we
have included in this paper a discussion of 't~Hooft's periodic
automaton model which gives at low energy scales rise to a genuine
quantum harmonic oscillator. In addition, such an automaton has a
close connection with our world crystal paradigm. Here we have
re-derived 't~Hooft's result in a new way. In contrast to 't~Hooft
derivation~\cite{tHoft2} we have matched the algebra of the
automaton variables with the $SU(1,1)$ algebra, and in contrast to
Ref.~\cite{JBV} we have worked with a different set of dynamical
variables. In the contraction limit (i.e., in the limit of many
lattice spacings --- low energy limit) we have recovered the
canonical $W(1)$-algebra and were able to identify the ``emergent"
harmonic oscillator variables.

There are several aspects of the double special relativity that are
worth noting in the connection with our generalized uncertainty
relation. Essentially, we have seen that the fundamental commutator in DSR as well as in our lattice GUP
goes to zero at Planck energy. For both models, the world should therefore be manifestly "classical" in the
Planck regime, a feature very different from the common believe.
This is a striking prediction supported by both models, although the lines of thought followed
in the two research paths are completely different and independent.
Moreover, this aspect presents also a strong resemblance with the results obtained
in the research line of ``deterministic" quantum mechanics.

%
%

\section{Acknowledgements}

One of us (P.J.) is grateful to G.~Vitiello for instigating
discussions. This work was partially supported by the Ministry of
Education of the Czech Republic (research plan MSM 6840770039), and
by the Deutsche Forschungsgemeinschaft under grant Kl256/47.
F.S. acknowledges financial support by the National Taiwan University under the
contract NSC 98-2811-M-002-086, and thanks ITP Freie Universit\"{a}t Berlin for
 warm hospitality.

\section*{Appendix:  Landauer principle}

Here we wish to provide an alternative derivation of the
mass-temperature formula (\ref{MT-})  based on Landauer's principle.
To this end we consider an ensemble of unpolarized photons that are
going to deliver to a micro black hole one single bit of information
per particle. In order to be sure that each photon delivers only one
bit of information
--- namely the information that it is there, somewhere inside in the black hole,
its position uncertainty must be ``maximal'', i.e., it should not be smaller than
Schwarzschild's radius $R_{S}$ as otherwise the photon would deliver
to black hole also extra bits of information concerning its
entry point (or better sector) on the horizon. At the same time its wavelength should
not be bigger than $R_{S}$, as otherwise the photon
would bounced off the black hole without getting trapped.
In this view, the position uncertainty of a photon in the ensemble must be of
order of Schwarzschild's radius $R_{S}$ i.e., $\Delta X_\epsilon \simeq \mu
R_{S}$. Factor $\mu$ is Bekenstein's deficit coefficient which ensures a correct
Hawking's formula in a continuum limit.

An extra bit of information added to the micro black hole will
increase its energy {\em at least} by amount $\Delta E_\epsilon$ so
that (cf. (\ref{26}))
\be \Delta X_\epsilon \Delta E_\epsilon  \ \simeq \ \frac{\hbar
c}{2}\left[1 - \frac{\epsilon^2}{2\hbar^2 c^2}\ \!
(\Delta E_\epsilon)^2 \right]. \label{A27} \ee
In the following we denote $\Delta E_\epsilon$ simply as $E_\epsilon$
to stress that $\Delta E_\epsilon$ an energy increas due to one photon.
With the explicit form for Planck's energy
\be \mathcal{E}_p \ =  \ \frac{\hbar c}{2\ell_p}  \ \approx \
0.61\cdot 10^{19} \ \!\mbox{GeV}\, , \label{Ape} \ee
the relation (\ref{A27}) can be cast to
%
\be \Delta X_\epsilon  \ \simeq \ \frac{\hbar c}{2 E_\epsilon} -
\frac{a^2\ell_p E_\epsilon}{8\mathcal{E}_p}\, . \label{A28} \ee

If we use further the fact that, $R_{S} =   \ell_p\, m $, where $m$
is the relative mass of the black hole in Planck units, i.e.,
$m=M/M_p$ ($M_p = \mathcal{E}_p/c^2$), we can rewrite  (\ref{A28})
as
\be 2m\mu \ \simeq  \ \frac{\mathcal{E}_{p}}{E_\epsilon} - \frac{a^2
E_\epsilon}{8\mathcal{E}_{p}}\,. \label{A29} \ee

According to the Landauer principle~\cite{landauer:61}, when a
single bit of information is erased (like in the black hole) the
amount of energy dissipated into environment is {\em at least} $k_B
T \ln 2$, where $k_B$ is Boltzmann's constant and $T$ is the
temperature of the erasing environment (in our case the micro black
hole). Since the liberated energy per bit of lost information can
not be grater the energy $E_\epsilon$ of the carrier photon   we
have that
\be E_\epsilon \ \simeq \  k_B T\, . \label{A111}
\label{energy} \ee
Relation (\ref{A111}) basically expresses equipartition law for an
unpolarized photon in the outgoing Hawking radiation. Defining the
Planck temperature $T_{p} = 2\mathcal{E}_{p}/k_B \approx
3\cdot10^{32}$~K, and measuring the temperature in terms of Planck
units as a relative temperature $\Theta = T/T_{p}$, we can rewrite
Eq.~(\ref{A29}) in the form
\be 2 m \  =  \   \frac{1}{2 \pi \Theta}
- 2\pi \zeta^2 \Theta \,.
\label{AMT-} \ee
where we identify $\zeta = a/(2\sqrt{2}\pi)$ and set $\mu = \pi$, in
order to agree with (\ref{MT-}) and with Hawking's formula (\ref{H})
in the continuum limit.


\end{document}